\documentstyle[seceq,preprint,epsbox]{jpsj}
\def\runtitle{On Phase Transition  of $\rm NH_{4}H_{2}PO_{4}$-Type Crystals}
\def\runauthor{Norihiro {\sc Ihara}, Shun-ichi {\sc Yoshida}, {\rm and} Koh {\sc{Wada}}}

\title{On Phase Transition  of $\rm NH_{4}H_{2}PO_{4}$-Type Crystals by Cluster Variation Method}
\author{Norihiro {\sc Ihara}\footnote{E-mail: ihara@statphys.sci.hokudai.ac.jp}, Shun-ichi {\sc Yoshida}, {\rm and} Koh {\sc{Wada}}}
\inst
{
Division of Physics, Graduate School of Science, Hokkaido University, Sapporo 060-0810
}
\recdate
{
\today
}

\abst
{
The Cluster Variation Method (CVM) is 
applied to the Ishibashi model for ammonium dihydrogen phosphate ($\rm NH_{4}H_{2}PO_{4}$) of a typical hydrogen bonded anti-ferroelectric  crystal. The 
staggered and the uniform susceptibility without hysteresis are calculated at 
equilibrium. On the other hand, by making use of the natural iteration method (NIM) for the CVM, hysteresis phenomena of uniform susceptibility versus temperature observed in experiments is well explained on 
the basis of local minimum in Landau type variational free energy. The polarization $P$ curves against the uniform field is also calculated.\\
}
\kword
{
 {$\rm NH_{4}H_{2}PO_{4}$ (ADP)}, anti-ferroelectrics, Ishibashi model, Cluster Variation Method (CVM), \\Natural Iteration Method (NIM), susceptibility 
}
\begin{document}
\sloppy
\maketitle
\newpage
\section{Introduction}
\ \ \ Recently, the pyroclore oxide  crystals A$_2$B$_2$O$_7$ (A=Ho, Dy, B=Sn, Ti) called the spin ice have drawn many attentions of researchers.\cite{Harris} The spin ice system has a typical geometrical frustration structure. The similar frustration 
structure is found in ammonium dihydrogen phosphate $\rm NH_{4}H_{2}PO_{4}$ (ADP). ADP is one of the hydrogen bonded crystals similar to well-known potassium 
dihydrogen phosphate ${\rm KH_{2}PO_{4}}$ (KDP). However, as a result of the 
replacement of potassium ion K$^{+}$ by ammonium ion NH$_4^{+}$, ADP undergoes 
the anti-ferroelectric phase transition while KDP is the typical ferroelectrics. Nagamiya\cite{Nagamiya} first suggested  a possibility of anti-ferroelectricity for ADP in the framework of the Slater's model for 
KDP\cite{Slater} in which the replacement of the first excited energy $\varepsilon_{0}$ due to hydrogen configuration around PO$_4$ by a negative value $-\varepsilon_{0}$ might well explain the experimental behaviors in ADP.  However, Ishibashi $et\  al.$\cite{Ishi2} pointed out that only taking negative value $-\varepsilon_{0}$ is not enough to realize the antiferroelectric phase transition 
because there are several proton configurations with the same 
energy as that proposed by Nagamiya due to geometrical frustration. In order to single out the observed crystal structure in ADP experiments,
Ishibashi  introduced the dipole-dipole interaction in Nagamiya's 
proposed model. Ishibashi\cite{Ishi1} further analyzed the extended model (hereafter Ishibashi model) in which it is 
possible for three or four protons to come closer to a PO$_4$ tetrahedron 
contrary to the ice rule. Here the ice rule demands: (1) each bond connecting 
the oxygen atoms in neighboring $\rm PO_{4}$ tetrahedra has always one and only one proton, and
(2) each $\rm PO_{4}$ tetrahedron can have only two protons near-by.
Namely the second rule of the ice rule is loosened in the Ishibashi model.  The model with this type of extension for the Slater's KDP theory is often called 
the Slater-Takagi model.\cite{Takagi}\par  
 In the present paper the Ishibashi model for ADP is reconsidered for the 
preliminary study of pyroclore crystals with geometrical frustration structure.
We calculate the staggered and the uniform susceptibility above and below 
the transition temperature in the Ishibashi model for the ADP-type crystal in 
the cactus approximation of the cluster variation method \cite{Kiku1} which is 
equivalent to the Slater's approximation for the 
KDP model.  Further, the hysteresis phenomena of uniform susceptibility versus 
temperature observed in experiments are successfully shown by making use of the natural iteration method (NIM) developed by Kikuchi\cite{Kiku2}. As the case of hysteresis phenomena depending upon the external electric field, the polarization $P$ curves against the uniform field is also calculated. \par 
In \S 2 we derive the variational free energy in the cactus approximation of the CVM.\cite{Wada2} In \S 3, from thermal equilibrium condition we obtain a self-consistent equation for polarization in a staggered electric field in order to find the properties of phase transition in the present system. After determining the tricritical point which stands for the boundary between the first and the second order 
phase transition, we calculate the staggered susceptibility and the 
antiferroelectric sublattice spontaneous polarization. In \S 4, after we 
calculate the uniform susceptibility of the present system in thermal equilibrium, we study hysteresis phenomena of uniform susceptibility versus temperature in order to compare with the experimental results. \S 5 is devoted to a summary. 

\section{Formulation}
\begin{figure}
\epsfile{file=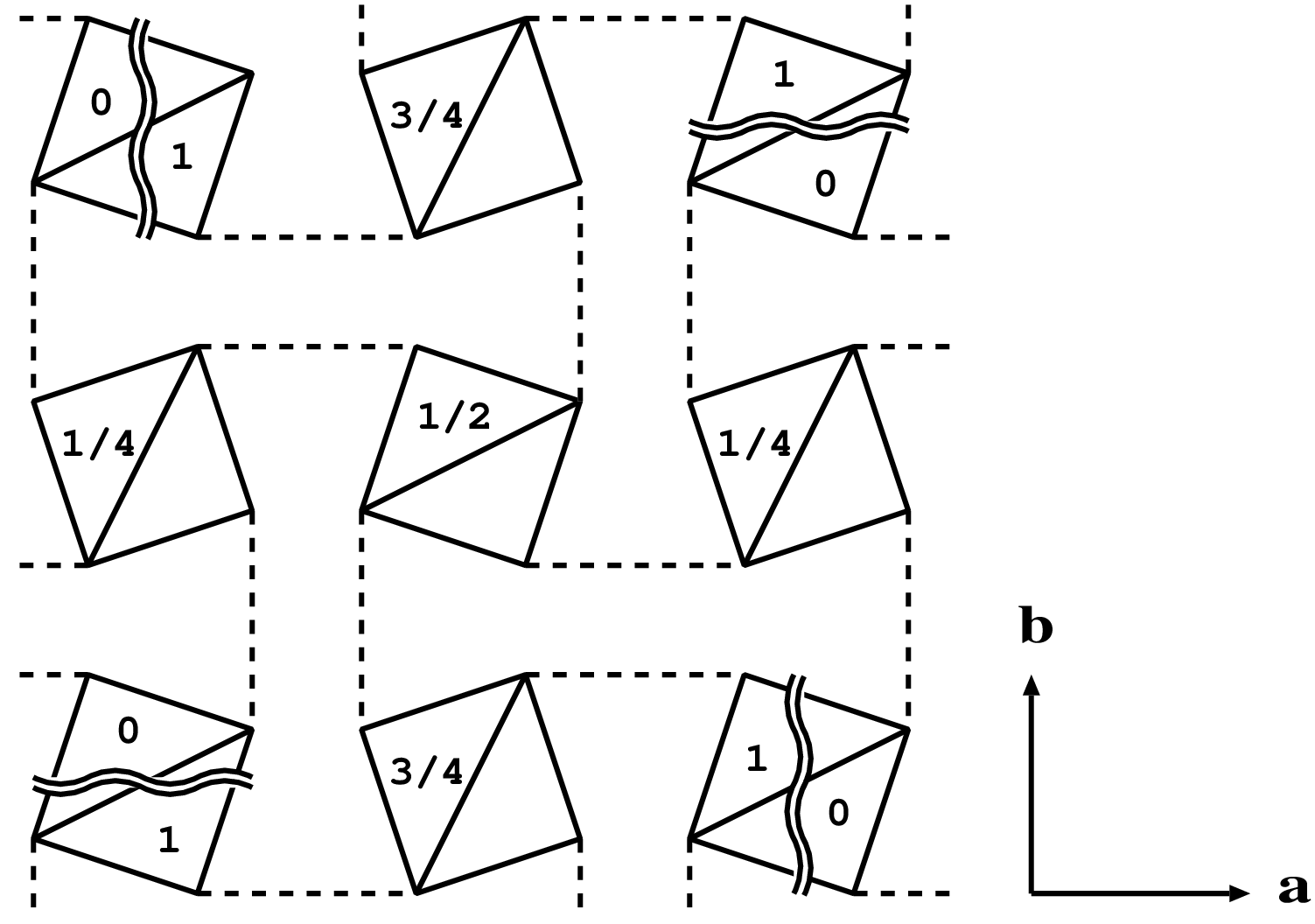}
\caption{The projection of atomic arrangement of ADP-type crystal on (001) plane. The number described in a $\rm PO_{4}$ tetrahedron represents the relative height of a $\rm PO_{4}$ tetrahedron.}
\label{cs}
\end{figure}

\begin{figure}
\epsfile{file=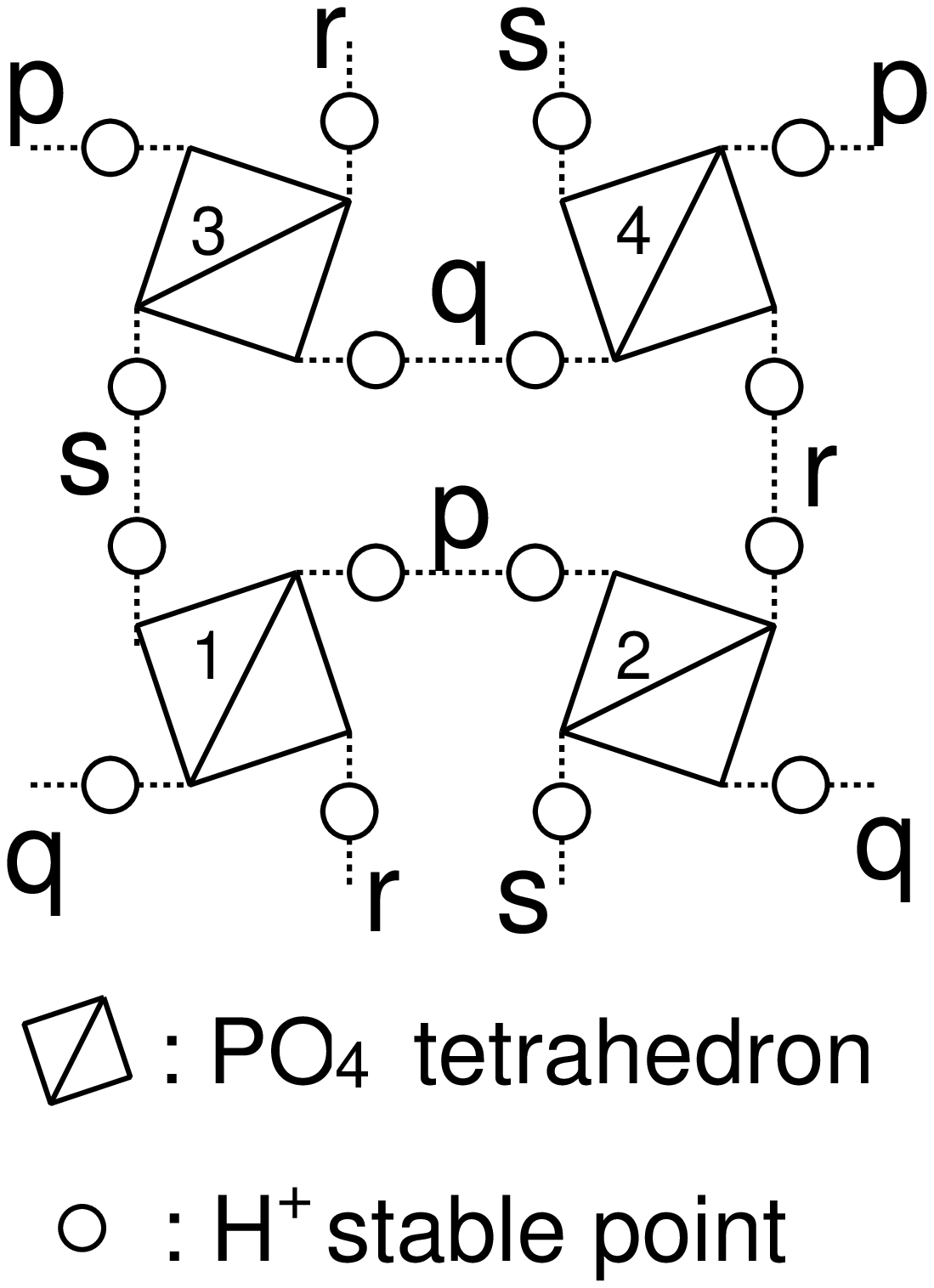}
\caption{Four sublatticies and hydrogen bonds connecting them. Two open circles on a hydrogen bond represent two stable points of a  proton.}
\label{fm}
\end{figure}

\begin{figure}
\epsfile{file=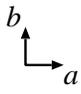,scale=0.8}
\caption{Energy, alloted dipole moment, and occurrence probability of proton configuration around $\rm PO_{4}$ tetrahedron belonging to $i$-th sublattice.}
\label{eap}
\end{figure}

\ \ \ Let us consider the system consisting of $2N$ protons around $N\ {\rm PO_{4}}$ tetrahedra in the $\rm ADP$-type crystal as shown in Fig. \ref{cs}. In order to formulate the present system in which the anti-ferroelectric phase 
transition along the $a$-axis takes place, we take a four 
sublattice model by Ishibashi\cite{Ishi1} as shown in Fig. \ref{fm}.  Here p, q, r and s denote non-equivalent hydrogen bonds on which a proton is located. When a proton on the hydrogen bond p is located close to the $i$-th sublattice, we 
denote the occurrence probability of such a proton configuration as $p_{i}$.
Next, let us assign the energy, the occurrence probability and the dipole moment along the $a$-axis to the protons configuration around PO$_4$ tetrahedron as shown in Fig. \ref{eap}.
We apply the cactus approximation of the CVM to the present system so as to find the variational free energy.  The cactus approximation of the CVM is equivalent to the Slater's theory  for KDP which takes account of the site of a proton in the double well potential along each O-O bond (hydrogen bond) between two 
nearest neighbor PO$_4$ tetrahedra and the correlation of four protons around 
each $\rm PO_{4}$ tetrahedron. In the following we call the occurrence probability of a 
proton configuration on the hydrogen bond as a bond (state) variable and that of four protons configuration around PO$_4$ tetrahedron as a four protons (state) variable.
The number of configurations of $L$ ensemble in the cactus 
approximation of the CVM is given by\cite{Kiku1}
\begin{eqnarray}
W=\prod_{i=1}^{2N} W_{bond}^{\langle i\rangle} \prod_{\langle ijkl\rangle}G^{\langle ijkl\rangle}_{tetra},
\end{eqnarray}
with
\begin{eqnarray}
 G^{\langle ijkl\rangle}_{tetra}=\frac{W^{\langle ijkl\rangle}_{tetra}}{W^{\langle i\rangle}_{bond}W^{\langle j\rangle}_{bond}
W^{\langle k\rangle}_{bond}W^{\langle l\rangle}_{bond}},
\end{eqnarray}
where, for example,  $W_{bond}^{\langle i\rangle}$ is the number of proton configurations on the $i$-th bond if the $i$-th bond is p bond between 1 and 2 sublattice:
\begin{eqnarray}
  W^{\langle i\rangle}_{bond} =\frac{L!}{(Lp_1)!(Lp_2)!},
\end{eqnarray}
and $G^{\langle ijkl\rangle}_{tetra}$ is the correlation number of protons for a PO$_4$ 
tetrahedron surrounded by protons on the i, j, k, l hydrogen bonds. And $W^{\langle ijkl\rangle}_{tetra}$ is the number of four protons configuration around PO$_4$ surrounded by protons on the i, j, k, l bonds. For example, referring to Fig. 3, for a PO$_4$ tetrahedron
belonging to the 1-st sublattice, $W^{\langle ijkl\rangle}_{tetra}$ is given by 
\begin{full}
\begin{equation}
W^{\langle ijkl\rangle}_{tetra}=\frac{L!}{[(Lc_{0}^{(1)})!]^2[(Lc_{2}^{(1)})!]^2
(La_{+}^{(1)})!(La_{-}^{(1)})![(La_{0}^{(1)})!]^2[(Ld_{+}^{(1)})!]^4[(Ld_{-}^{(1)})!]^4},
\end{equation}
\end{full}
where the superscript $(1)$ denotes the number of sublattice.
By utilizing Stirling's formula and assuming the homogeneity in each  sublattice, the entropy of the present four sublattice model  becomes
\begin{eqnarray}
\frac{S}{k_{\rm B}}&=&\frac{1}{L}\log W\nonumber\\
&=&\frac{N}{4}\sum^{4}_{i=1}\bigl[p_{i}\ln p_{i}+q_{i}\ln q_{i}+r_{i}\ln r_{i}+s_{i}\ln s_{i}\nonumber\\
&&-\left(a^{(i)}_{+}\ln a^{(i)}_{+}+a^{(i)}_{-}\ln a^{(i)}_{-}+4d^{(i)}_{+}\ln d^{(i)}_{+}\right.\nonumber\\
&&+4d^{(i)}_{-}\ln d^{(i)}_{-}+2a^{(i)}_{0}\ln a^{(i)}_{0}\nonumber\\
&&\left. +2c^{(i)}_{0}\ln c^{(i)}_{0}+2c^{(i)}_{2}\ln c^{(i)}_{2}\right)\bigr]. \label{entropy}
\end{eqnarray}
Further, the electric polarization of $i$-th 
sublattice per PO$_4$ along the $a$-axis is  defined by 
\begin{eqnarray}
 \mu_a P^{(i)} =\mu_a [(a_{+}^{(i)}-a_{-}^{(i)})+2(d_{+}^{(i)}-d_{-}^{(i)})]\quad\quad (i=1\sim 4),\label{polar}\nonumber\\
\end{eqnarray}
where $\mu_a$ is the dipole moment along the $a$-axis. The proton 
configuration energy $U$ per system is given by 
\begin{eqnarray}
U&=&\frac{N}{4}\sum_{i=1}^4[2\varepsilon_0c_0^{(i)}
+4\varepsilon_1(d_{+}^{(i)}+d_{-}^{(i)})+2\varepsilon_2c_2^{(i)}]\nonumber\\
&&+\frac{N}{4}\lambda \mu_a^2(P^{(1)}+P^{(2)})(P^{(3)}+P^{(4)})\label{uenergy}\\
&&-\frac{N}{4}[\mu_aE(P^{(1)}+P^{(2)})+\mu_aE'(P^{(3)}+P^{(4)})],\nonumber
\end{eqnarray}
where the first line represents the protons configuration energy around $\rm PO_{4}$ tetrahedra, the second line denotes effectively the long range dipole-dipole interaction energy to induce an anti-ferroelectric structure along the $a$-axis and the last line is the energy due to external 
electric field. In the following we call the case of $E'=-E$  a staggered 
electric field 
and  the case of $E'=E$ a homogeneous electric field. Here it should be  noted 
that bond variables are always expressed in terms of four protons variables:
\begin{eqnarray}
2p_1=2q_2&=&c_0^{(1)}+c_2^{(1)}+a_0^{(1)}+a_{+}^{(1)}+3d_{+}^{(1)}
+d_{-}^{(1) }\nonumber\\&&
+c_0^{(2)}+c_2^{(2)}+a_0^{(2)}+a_{+}^{(2)}+3d_{+}^{(2)}
+d_{-}^{(2) },\nonumber\\
2p_2=2q_1&=&c_0^{(1)}+c_2^{(1)}+a_0^{(1)}+a_{-}^{(1)}+d_{+}^{(1)}+3d_{-}^{(1)}
\nonumber\\&&+c_0^{(2)}+c_2^{(2)}+a_0^{(2)}+a_{-}^{(2)}+d_{+}^{(2)}+3d_{-}^{(2)},\nonumber\\
2r_1=2s_3&=&c_0^{(1)}+c_2^{(1)}+a_0^{(1)}+a_{+}^{(1)}+3d_{+}^{(1)}+d_{-}^{(1)}
\nonumber\\&&+c_0^{(3)}+c_2^{(3)}+a_0^{(3)}+a_{-}^{(3)}+d_{+}^{(3)}+3d_{-}^{(3)},\nonumber\\
2r_3=2s_1&=&c_0^{(1)}+c_2^{(1)}+a_0^{(1)}+a_{-}^{(1)}+d_{+}^{(1)}+3d_{-}^{(1)}
\nonumber\\&&+c_0^{(3)}+c_2^{(3)}+a_0^{(3)}+a_{+}^{(3)}+3d_{+}^{(3)}+d_{-}^{(3)},\nonumber\\
2p_3=2q_4&=&c_0^{(3)}+c_2^{(3)}+a_0^{(3)}+a_{-}^{(3)}+d_{+}^{(3)}+3d_{-}^{(3)}
\nonumber\\&&+c_0^{(4)}+c_2^{(4)}+a_0^{(4)}+a_{-}^{(4)}+d_{+}^{(4)}+3d_{-}^{(4)},\nonumber\\
2p_4=2q_3&=&c_0^{(3)}+c_2^{(3)}+a_0^{(3)}+a_{+}^{(3)}+3d_{+}^{(3)}+d_{-}^{(3)}
\nonumber\\&&+c_0^{(4)}+c_2^{(4)}+a_0^{(4)}+a_{+}^{(4)}+3d_{+}^{(4)}+d_{-}^{(4)},\nonumber\\
2r_2=2s_4&=&c_0^{(2)}+c_2^{(2)}+a_0^{(2)}+a_{+}^{(2)}+3d_{+}^{(2)}+d_{-}^{(2)}
\nonumber\\&&+c_0^{(4)}+c_2^{(4)}+a_0^{(4)}+a_{-}^{(4)}+d_{+}^{(4)}+3d_{-}^{(4)},\nonumber\\
2r_4=2s_2&=&c_0^{(2)}+c_2^{(2)}+a_0^{(2)}+a_{-}^{(2)}+d_{+}^{(2)}+3d_{-}^{(2)}
\nonumber\\&&+c_0^{(4)}+c_2^{(4)}+a_0^{(4)}+a_{+}^{(4)}+3d_{+}^{(4)}+d_{-}^{(4)}.\nonumber\\
\end{eqnarray}
There are also normalization relations among four protons state variables:
\begin{eqnarray}
&&a^{(i)}_{+}+a^{(i)}_{-}+4(d^{(i)}_{+}+d^{(i)}_{-})+2a^{(i)}_{0}+2c^{(i)}_{0}+2c^{(i)}_{2}=1 \label{nomal1}\nonumber\\
&&\hspace{5.5cm}(i=1\sim 4),
\end{eqnarray}
where the superscript $(i)$ denotes the sublattice number. 
Finally, by combining eq.(\ref{entropy}) and eq.(\ref{uenergy}) the variational free energy $G$ in the cactus approximation is obtained by 
\begin{eqnarray}
G =U-TS.
\end{eqnarray}
\section{Response to Staggered Electric Field}
\ \ \ Let us calculate the staggered susceptibility to study the properties of phase transition of the system. The staggered field is applied so as to induce the anti-ferroelectric order. Since the anti-ferroelectric structure is assumed to occur along the $a$-axis, the electric field $E'=-E$ is applied to the sublattice 3 and 4 in addition of $E$ to the sublattice 1 and 2. Since the sublattice 3 and 4 
are equivalent to 1 and 2 except having a sublattice polarization in the opposite 
direction, the present system is reduced to the one sublattice problem and we have 
following relations 
\begin{eqnarray}
a_0^{(i)}&=&a_0, \quad c_0^{(i)}=c_0,\quad c_2^{(i)}=c_2 \quad(i=1\sim 4),\nonumber\\
a_{\pm}^{(1)}&=&a_{\pm}^{(2)}=a_{\mp}^{(3)}=a_{\mp}^{(4)}=a_{\pm}, \nonumber\\
d_{\pm}^{(1)}&=&d_{\pm}^{(2)}=d_{\mp}^{(3)}=d_{\mp}^{(4)}=d_{\pm}, \nonumber\\
p_1&=&p_3=q_2=q_4=r_1=r_2=s_3=s_4\nonumber\\&=&c_0+c_2+a_0+a_{+}+3d_{+}+d_{-},\nonumber\\
p_2&=&p_4=q_1=q_3=r_3=r_4=s_1=s_2\nonumber\\&=&c_0+c_2+a_0+a_{-}+d_{+}+3d_{-}.
\end{eqnarray}
And the sublattice polarization conjugate to the staggered field is now 
given by 
\begin{eqnarray}
P\equiv P^{(1)}&=&P^{(2)}=-P^{(3)}=-P^{(4)}\nonumber\\&=&a_{+}-a_{-}+2(d_{+}-d_{-})=p_1-p_2.
\label{smoment}
\end{eqnarray}
The  free energy $G$ of the system takes a form:
\begin{eqnarray}
\frac{G}{Nk_{\rm B}T}&=&\left[\frac{2\varepsilon_0}{k_{\rm B}T}c_0+\frac{4\varepsilon_1}{k_{\rm B}T}(d_{+}+d_{-})+\frac{2\varepsilon_2}{k_{\rm B}T}c_2\right] -\frac{\lambda\mu_a^2}{k_{\rm B}T}P^2\nonumber\\
&&\hspace{-0.5cm}-\frac{\mu_aEP}{k_{\rm B}T}
+[-2(\frac{1+P}{2}\ln{\frac{1+P}{2}}+\frac{1-P}{2}\ln{\frac{1-P}{2}})\nonumber\\&&\hspace{-0.5cm}
+(a_{+}\ln{a_{+} }+a_{-}\ln{a_{-} }+2a_{0}\ln{a_{0} }\nonumber\\
&&\hspace{-0.5cm}+2c_{0}\ln{c_{0} }+2c_{2}\ln{c_{2} }+4d_{+}\ln{d_{+} }
+4d_{-}\ln{d_{-} })]\label{sfree}\nonumber\\
&&\hspace{-0.5cm}+\frac{\gamma}{k_{\rm B}T}\left[1-(a_{+}+a_{-}+4(d_{+}+d_{-})+2a_{0}+2c_{0}+2c_{2})\right],
\end{eqnarray} 
where $\gamma$ in the last line is the Lagrange multiplier to make all the state variables $a_{+}, a_{-}, a_{0}, d_{+}, d_{-}, c_{0}, c_{2}$ independent. Under the staggered 
electric field the thermal equilibrium state is obtained 
from the minimum condition  of the free energy: $\frac{\partial G}{\partial a_{+}}=\frac{\partial G}{\partial a_{-}}=\frac{\partial G}{\partial d_{+}}=\frac{\partial G}{\partial d_{-}}=\frac{\partial G}{\partial a_{0}}=\frac{\partial G}{\partial 
c_{0}}=\frac{\partial G}{\partial c_{2}}=0$. The state variables are solved in terms of $a_0$, the electric polarization $P$ and the staggered field $E$ as follows:
\begin{eqnarray}
c_0&=&\eta_0a_0,\quad\quad c_2=\eta_2a_0, \nonumber\\
a_{+}&=&a_0A_Ph^2, \quad\quad a_{-}=a_0(A_Ph^{2})^{-1},\label{seqv}\\
d_{+}&=&a_0\eta_1{A_P}^{\frac{1}{2}}h,\quad\quad d_{-}=a_0\eta_1{A_P}^{-\frac{1}{2}}h^{-1},\nonumber
\end{eqnarray}
where $\eta_0=\exp{(-\varepsilon_0/k_{\rm B}T)}$, $\eta_1=\exp{(-\varepsilon_1/k_{\rm B}T)}$, $\eta_2=\exp{(-\varepsilon_2/k_{\rm B}T)}$, 
$h=\exp{\frac{\mu_aE}{2k_{\rm B}T}}$ 
 and an abbreviation is defined as 
\begin{eqnarray}
A_P =\frac{1+P}{1-P}\exp{(2DP)}\quad (D\equiv\frac{\lambda\mu_a^2}{k_{\rm B}T}).
\end{eqnarray}
Further, $a_0$ is determined by a normalization condition as
\begin{eqnarray}
a_{0}&=&\left[2+2\eta_{0}+2\eta_{2}+A_{P}h^2+{A_{P}}^{-1}h^{-2}\nonumber\right.\\
&&\left.+4\eta_{1}({A_{P}}^{\frac{1}{2}}h+{A_P}^{-\frac{1}{2}}h^{-1})\right]^{-1}.
\label{seqa}
\end{eqnarray}
Substituting eq.(\ref{seqv}) and eq.(\ref{seqa}) into the variational free 
energy of eq.($\ref{sfree}$), the equilibrium free energy $G_{\rm e}$ under the staggered field is given by
\begin{eqnarray}
\frac{G_{\rm e}}{Nk_{\rm B}T} =\ln{\frac{4a_{0}}{(1-P^2)}}+\frac{\lambda}{k_{\rm B}T}P^2.
\label{eqfree}
\end{eqnarray}
Without staggered field this expression has been obtained by Ishibashi.\cite{Ishi1} The sublattice polarization $P$ under the staggered field is obtained as a 
self-consistent equation from eq.({\ref{smoment}}):
\begin{eqnarray}
P =  a_0\left[A_Ph^2-{A_P^{-1}h^{-2}}+2\eta_1({A_P}^{\frac{1}{2}}h-{{A_P}}^{-\frac{1}{2}}h^{-1})
\right].\label{self}\nonumber\\
\end{eqnarray}
The same self-consistent equation for $P$  is also obtained by the thermodynamic relation $N\mu_aP = -\frac{1}{k_{\rm B}T}\frac{\partial G}{\partial E}$.\par
\begin{figure}
\epsfile{file=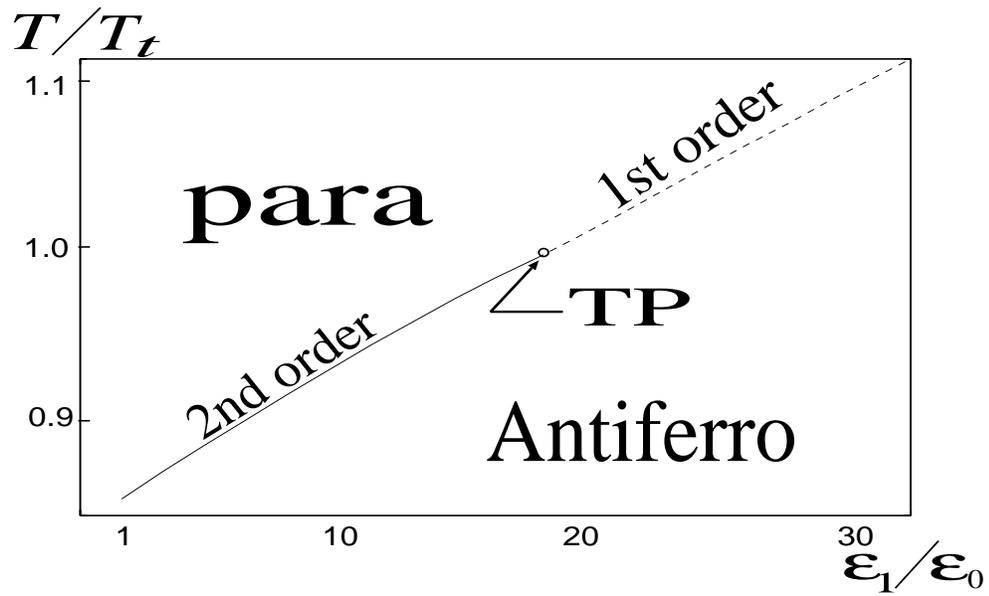,height=8cm,width=13cm}
\caption{The phase diagram in $\varepsilon_{1}-T$ space with $\varepsilon_{0}=6k_{\rm B}, \varepsilon_{2}=1000k_{\rm B}, \lambda=232k_{\rm B}$. The dotted line and solid line represents the first order  and the second order phase transition temperature, respectively, and the boundary circle represents the tricritical point(TP).}
\label{etc}
\end{figure}

\begin{figure}
\epsfile{file=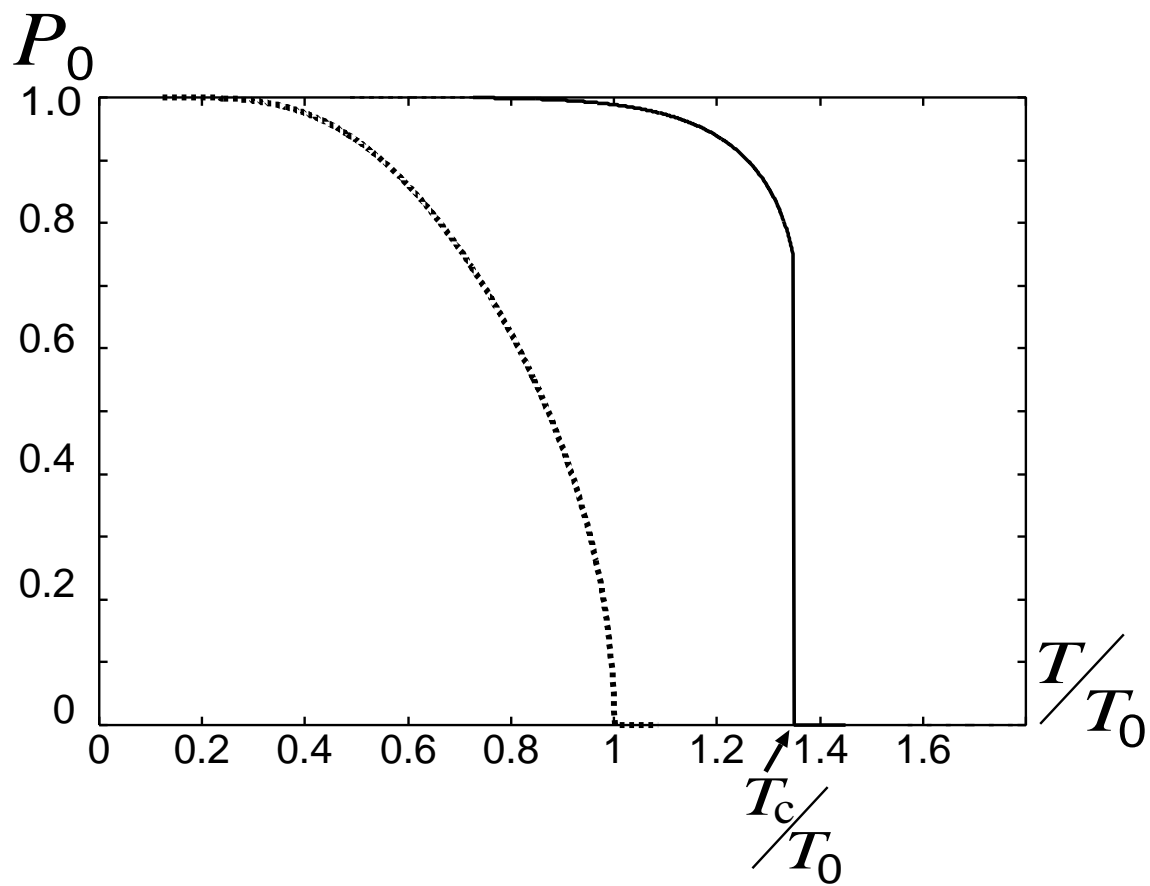,scale=0.8}
\caption{The temperature dependence of antiferroelectric spontaneous polarization $P_{0}$. The dotted line ($\varepsilon_{1}=10k_{\rm B}$:second order) and the solid line ($\varepsilon_{1}=200k_{\rm B}$:first order) are shown.}
\label{figop}
\end{figure}

\begin{figure}
\epsfile{file=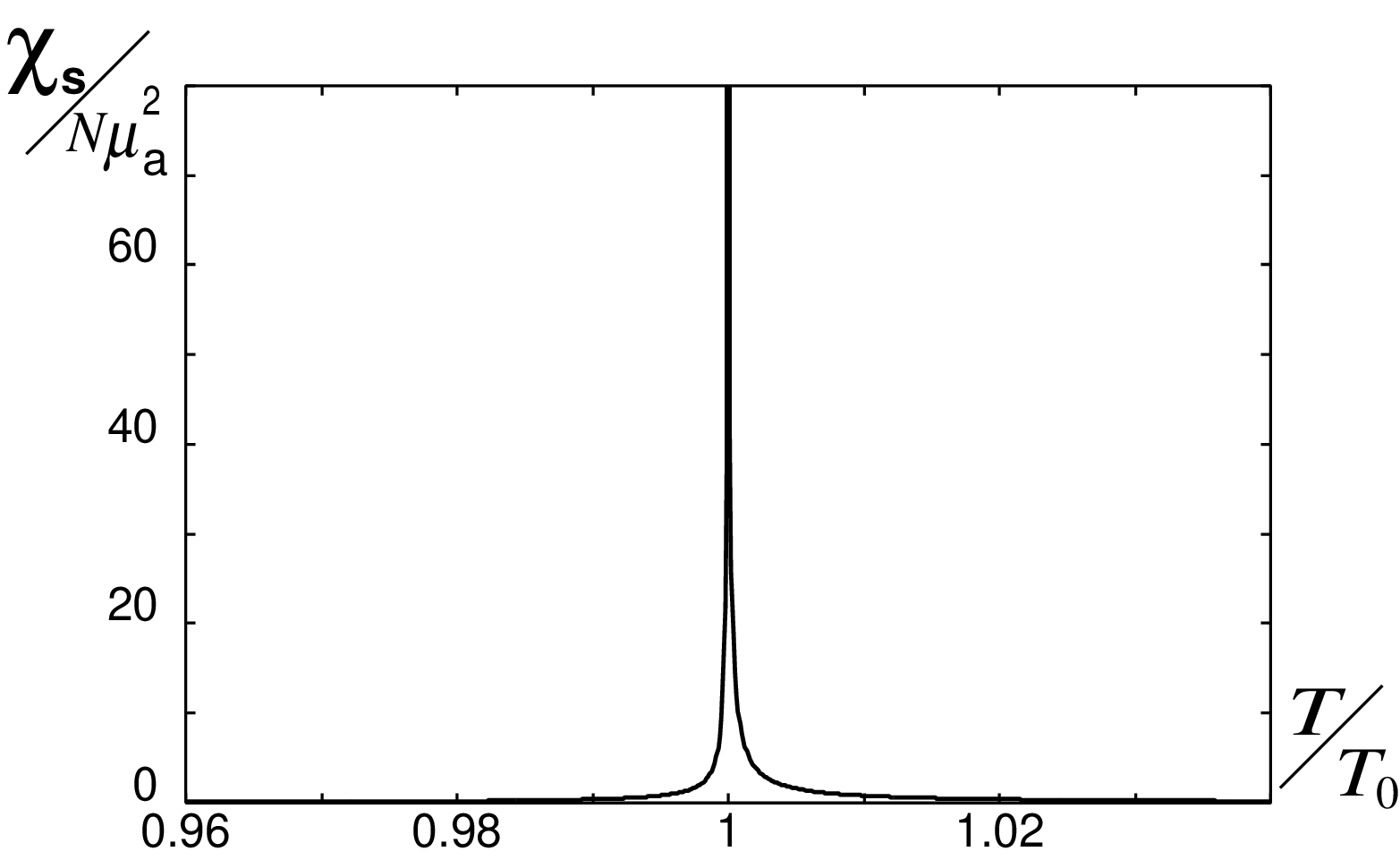,scale=0.4}
\epsfile{file=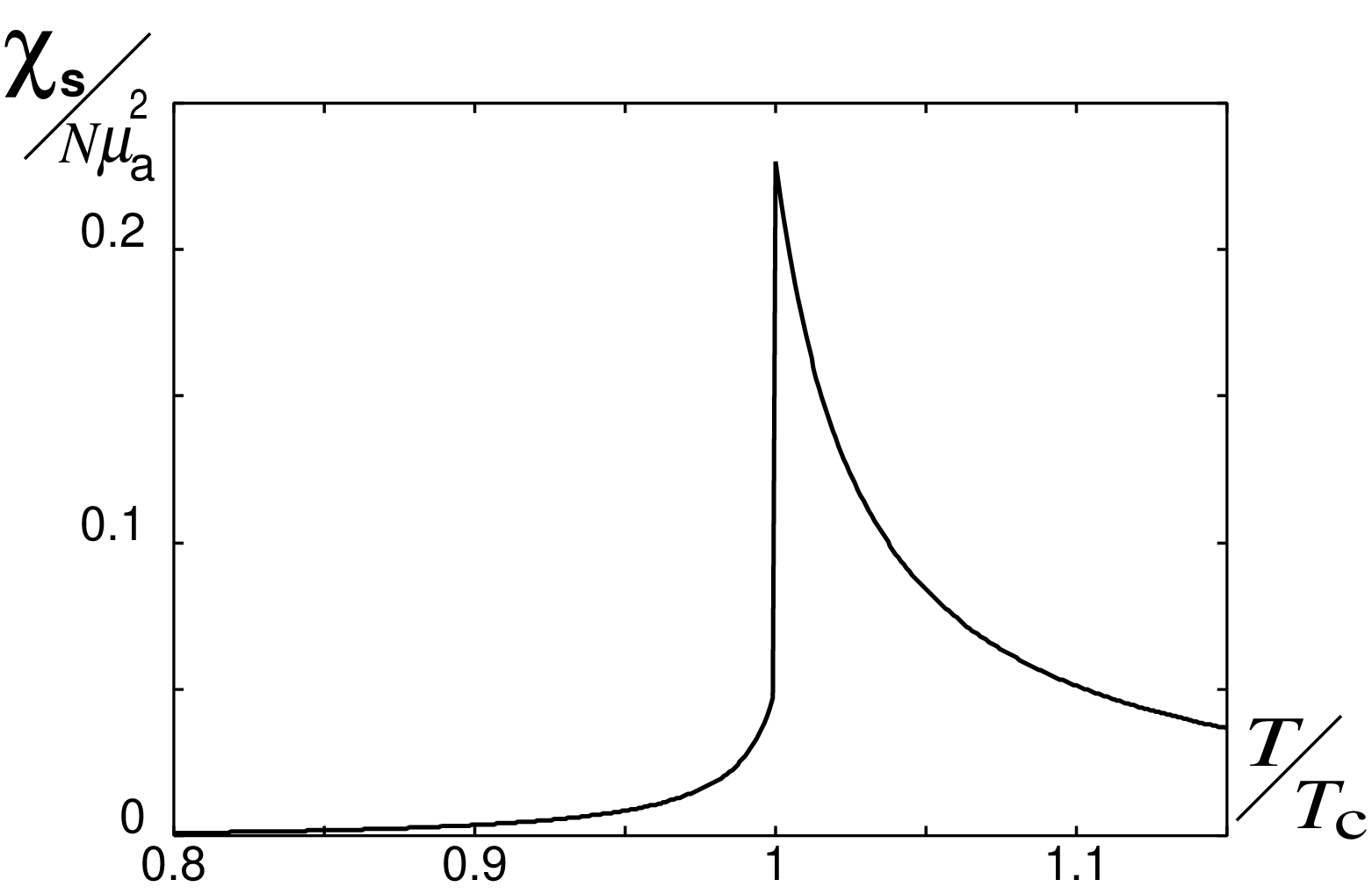,scale=0.4}
\caption{The temperature dependence of staggered susceptibility $\chi_{s}$. The left figure is for the second order case of $\varepsilon_{1}=10k_{\rm B}$ and the right one for the first order case of $\varepsilon_{1}=200k_{\rm B}$.}
\label{ss}
\end{figure}

 
In order to investigate the phase transition properties we expand the above 
equation (\ref{self}) up to the 3-rd order of polarization $P$ and to the linear order of staggered field $E$ in the neighborhood of the transition temperature: 
\begin{eqnarray}
A_2(T)P+A_4(T)P^3+\cdots -4(1+\eta_1)\frac{\mu_aE}{k_{\rm B}T}=0,\nonumber\\
\end{eqnarray}
where  $A_2(T)$ and $A_4(T)$ are given by 
\begin{eqnarray}
A_2(T)&=&2\eta_0+4\eta_1+2\eta_2-4(1+\eta_1)D,\nonumber\\
A_4(T)&=&-4D^2-\frac{8D^3}{3}+2\eta_1(1+3D+D^2-\frac{D^3}{3}). \nonumber\\
\end{eqnarray}
From the view point of Landau's phase transition theory the order of the phase 
transition  is classified as follows. (i) The phase transition undergoes the 
second order transition at $T_0$ if $A_2(T_0)=0$ and $A_4(T_0)>0$. (ii) The 
phase transition undergoes the first order transition at $T_C (>T_0)$ if $A_2(T_0)=0$ and $A_4(T_0)<0$. The boundary between the first and the second order 
transition 
is called the tricritical point $T_t(=T_0)$ when $A_2(T_0)=0$ and $A_4(T_0)=0$. The phase diagram in the $\varepsilon_1-T$ space is shown in Fig. \ref{etc}. 
Hereafter the parameters $\varepsilon_0=6k_{\rm B}, \varepsilon_2=1000k_{\rm B}, \lambda=232k_{\rm B}$ are taken. It can be 
seen that the 
energy parameter $\varepsilon_1$ representing HPO$_4$ and H$_3$PO$_4$ controls  the 
order of phase transition. As $\varepsilon_1/\varepsilon_0$ is increased, the second 
order phase transition for small $\varepsilon_1/\varepsilon_0$ changes into the first order transition. 
In any case the spontaneous sublattice polarization $P_0$ of the 
anti-ferroelectric state is given by the self-consistent equation (\ref{self}) without electric field: 
\begin{eqnarray}
P_0 =\hat{a}_0[A_{P_0}-{A_{P_0}}^{-1}+2\eta_1({A_{P_0}}^{\frac{1}{2}}
-{A_{P_0}}^{-\frac{1}{2}})],
\label{P_0}\nonumber\\
\end{eqnarray}
where $\hat{a}_0$ is a thermal equilibrium value of $a_0$ in eq.(\ref{seqa}) without external field. We solve eq.($\ref{P_0}$) numerically and show the spontaneous polarization $P_0$ versus temperature in Fig. \ref{figop}. On the other side, the staggered 
susceptibility $\chi_{\rm s} $ of the system is obtained from eq.(\ref{self}) as the 
linear response $\Delta P$ from $P_0$ induced by the staggered field $E$
\begin{eqnarray}
\chi_{\rm s}(T)&=&\lim_{E\to 0}N\mu_a\frac{\Delta P}{E}
=\frac{N\mu_a^2}{k_{\rm B}T}\frac{1}{X^{-1}-(\frac{1}{1-P_0^2}+D)}
\label{stsus}\nonumber\\
\end{eqnarray}
with
\begin{eqnarray}
X&=&2\hat{a_0}\left[A_{P_0}+{A_{P_0}}^{-1}+\eta_1({A_{P_0}}^{\frac{1}{2}}+{A_{P_0}}^{-\frac{1}{2}})\right]-2P_0^2.\nonumber
\end{eqnarray}
Especially, in the paraelectric phase the staggered susceptibility is simplified into
\begin{eqnarray}
\chi_{\rm s}(T) =\frac{N\mu_a^2}{k_{\rm B}T}\frac{2(1+\eta_1)}{\eta_0+2\eta_1+\eta_2-2(1+\eta_1)D}.
\end{eqnarray}
The staggered susceptibility $\chi_{\rm s}(T)$ is shown in Fig. \ref{ss} for each case of 
the second and the first order transition. It is noteworthy that when we choose $D=0$, the staggered susceptibility never diverges.  Without a finite dipole-dipole interaction parameter $\lambda$, we would have the paraelectric state down 
to the zero temperature owing to geometrical frustration structure of the 
system.

\section{Response to Uniform Electric Field}
\begin{figure}
\epsfile{file=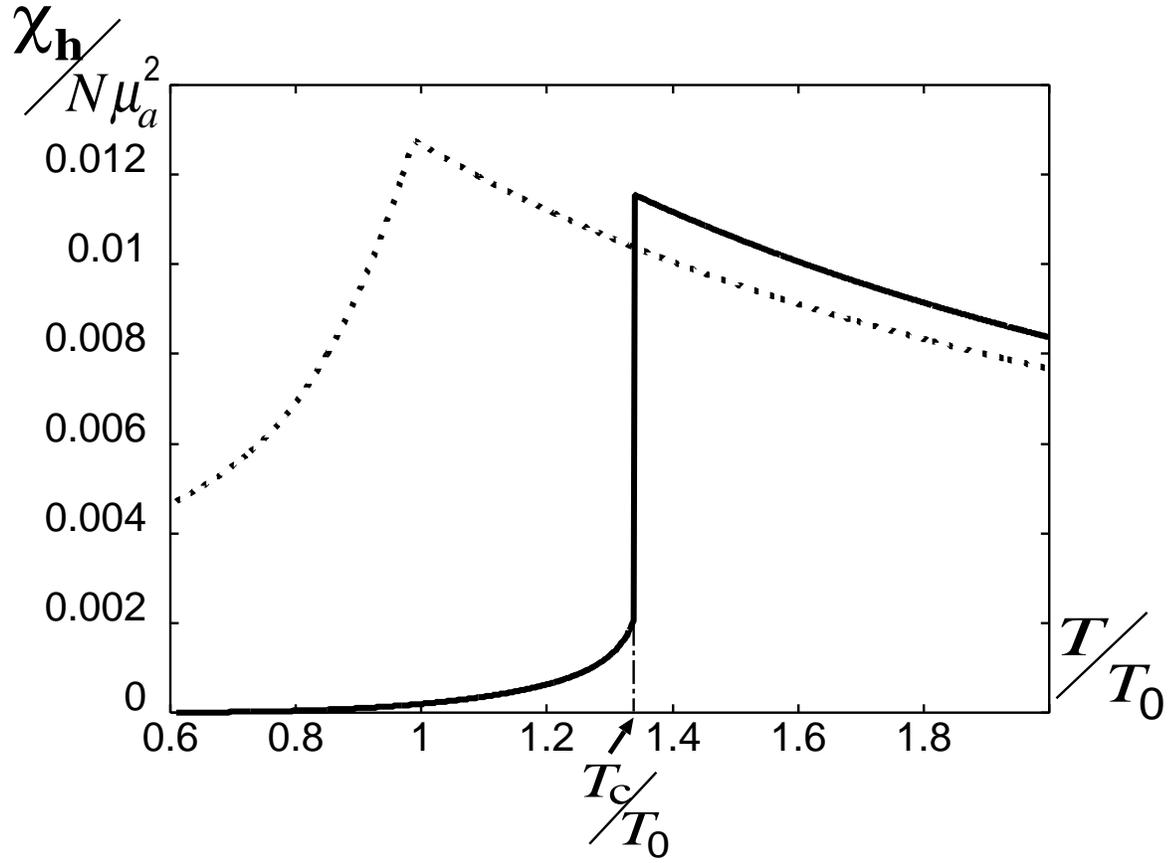,scale=0.8}
\caption{The temperature dependence of uniform susceptibility $\chi_{\rm h}$. The dotted line ($\varepsilon_{1}=10k_{\rm B}$:second order) and the solid line ($\varepsilon_{1}=200k_{\rm B}$:first order) are shown.}
\label{unieq}
\end{figure}

\ \ \ Let us study the response of the present system to a uniform electric 
field along the $a$-axis. The same external electric field $E'=E$ is applied to the sublattice 3 and 4 as well as to the sublattice 1 and 2. Contrary to the case of staggered electric field, in the anti-ferroelectric state the sublattice 
1 and 2 are non-equivalent to the sublattice 3 and 4 in response to the uniform 
electric field. The present system is reduced to the two sublattice problem and we have following relations: 
\begin{eqnarray}
a_0^{(i)}&=&a_0,\quad c_0^{(i)}=c_0,\quad c_2^{(i)}=c_2\quad (i=1\sim 4),
\nonumber\\
a_{\pm}^{(1)}&=&a_{\pm}^{(2)}=a_{\pm},\quad\quad a_{\pm}^{(3)}=a_{\pm}^{(4)}=a'_{\mp}, \nonumber\\
d_{\pm}^{(1)}&=&d_{\pm}^{(2)}=d_{\pm},\quad\quad d_{\pm}^{(3)}=d_{\pm}^{(4)}=d'_{\mp},\label{defuni} \\
p_1&=&q_2=c_0+c_2+a_0+a_{+}+3d_{+}+d_{-},\nonumber\\p_2&=&q_1=c_0+c_2+a_0+a_{-}+d_{+}+3d_{-},
\nonumber\\
p_3&=&q_4=c_0+c_2+a_0+a'_{+}+3d'_{+}+d'_{-}, \nonumber\\p_4&=&q_3=c_0+c_2+a_0+a'_{-}+d'_{+}+3d'_{-},
\nonumber\\
2r_1&=&2s_3=2r_2=2s_4\nonumber\\
&&\hspace{-1cm}=
2c_0+2c_2+2a_0+a_{+}+a'_{+}+3(d_{+}+d'_{+})+(d_{-}+d'_{-}),\nonumber\\
2r_3&=&2s_1=2r_4=2s_2\nonumber\\
&&\hspace{-1cm}=2c_0+2c_2+2a_0+a_{-}+a'_{-}+(d_{+}+d'_{+})+3(d_{-}+d'_{-}).\nonumber
\end{eqnarray}
And the sublattice electric polarizations along the $a$-axis are defined as
\begin{eqnarray}
P&\equiv &P^{(1)}=P^{(2)} =a_{+}-a_{-}+2(d_{+}-d_{-})=p_1-p_2,\label{spolar}\nonumber\\
P'&\equiv&-P^{(3)}=-P^{(4)}=a_{+}'-a_{-}'+2(d_{+}'-d_{-}')=p_3-p_4.\nonumber\\
\label{hmoment}
\end{eqnarray}
The variational free energy $G$ in the uniform field is rewritten as 
\begin{eqnarray}
\frac{G}{Nk_{\rm B}T}&=&2[\frac{\varepsilon_0 c_0}{k_{\rm B}T}+\frac{\varepsilon_1}{k_{\rm B}T}(d_{+}+d_{-}+d_{+}'+d_{-}')+\frac{\varepsilon_2 c_2}{k_{\rm B}T}]\nonumber\\ 
&&\hspace{-1.3cm}-\frac{\lambda\mu_a^2}{k_{\rm B}T}PP'-\frac{\mu_aE}{2k_{\rm B}T}(P-P')
\nonumber\\
&&\hspace{-1.3cm}-\frac{1}{2}[(p_1\ln{p_1}+p_2\ln{p_2})+2(r_1\ln{r_1}+r_3\ln{r_3})
\nonumber\\
&&\hspace{-1.3cm}+(p_3\ln{p_3}+p_4\ln{p_4})]\nonumber\\
&&\hspace{-1.3cm}+\frac{1}{2}[a_{+}\ln{a_{+} }+a_{-}\ln{a_{-} }+a_{+}'\ln{a_{+}' }+a_{-}'\ln{a_{-}' }\nonumber\\
&&\hspace{-1.3cm}+4a_{0}\ln{a_{0} }+4c_{0}\ln{c_{0} }+4c_{2}\ln{c_{2} }\nonumber\\
&&\hspace{-1.3cm}+4(d_{+}\ln{d_{+} }+d_{-}\ln{d_{-} })
+4(d_{+}'\ln{d_{+}' }+d_{-}'\ln{d_{-}' })]\nonumber\\
&&\hspace{-1.3cm}+\frac{\gamma_1}{k_{\rm B}T}[1-(a_{+}+a_{-}+4(d_{+}+d_{-})+2a_{0}+2c_{0}+2c_{2})]
\nonumber\\
&&\hspace{-1.3cm}+\frac{\gamma_2}{k_{\rm B}T}[1-(a_{+}'+a_{-}'+4(d_{+}'+d_{-}')+2a_{0}+2c_{0}+2c_{2})],\label{hfree}\nonumber\\
\end{eqnarray} 
where $\gamma_1$ and $\gamma_2$ are Lagrange multipliers which are determined by the normalization relations (\ref{nomal1}) with the help of eq.(\ref{defuni}). 
The thermal equilibrium is determined by the minimum condition of the 
free energy with respect to independent variables: 
$\frac{\partial G}{\partial a_{+}}=\frac{\partial G}{\partial a_{-}}
=\frac{\partial G}{\partial d_{+}}=\frac{\partial G}{\partial d_{-}}
=\frac{\partial G}{\partial a'_{+}}=\frac{\partial G}{\partial a'_{-}}
=\frac{\partial G}{\partial d'_{+}}=\frac{\partial G}{\partial d'_{-}}
=\frac{\partial G}{\partial a_{0}}=\frac{\partial G}{\partial c_{0}}
=\frac{\partial G}{\partial c_{2}}=0$.  
These relations are regarded as a set of equations of 
state in the present system under a homogeneous electric field. In Appendix A  it is shown that all the independent variables are expressed in terms of sublattice polarizations $P$ and $P'$ under the external electric field $E$. Since the linear response of sublattice polarization to the uniform field $E$ is written as $\Delta 
P =P-P_0 $ for the 1 and 2 sublattices and $\Delta P'=P'-P_0$ for the 3 and 4 sublattices,   
the uniform susceptibility $\chi_{\rm h}$ is defined as 
\begin{eqnarray}
\chi_{\rm h} =\lim_{E\to 0}\frac{N}{2}\mu_a^2\frac{\Delta P -\Delta P'}{E}.
\end{eqnarray}
Since concrete calculations are complicated and tedious, details of the 
calculation and the final result are given in Appendix A.  Here we mention only the homogeneous susceptibility of the system in the para-electric phase
\begin{eqnarray}
\chi_{\rm h}=\frac{N\mu^2_{a}}{k_{\rm B}T}\frac{4+4\eta_{1}}{1+\eta_{0}+3\eta_{1}+\eta_{2}+\frac{D}{2}(1+\eta_{1})}.\label{unip}
\end{eqnarray}
The numerical results of  the uniform susceptibility $\chi_{\rm h}$ against 
temperature are shown in Fig. \ref{unieq} for each case of the second order and the first order transition.
\par
\begin{figure}
\epsfile{file=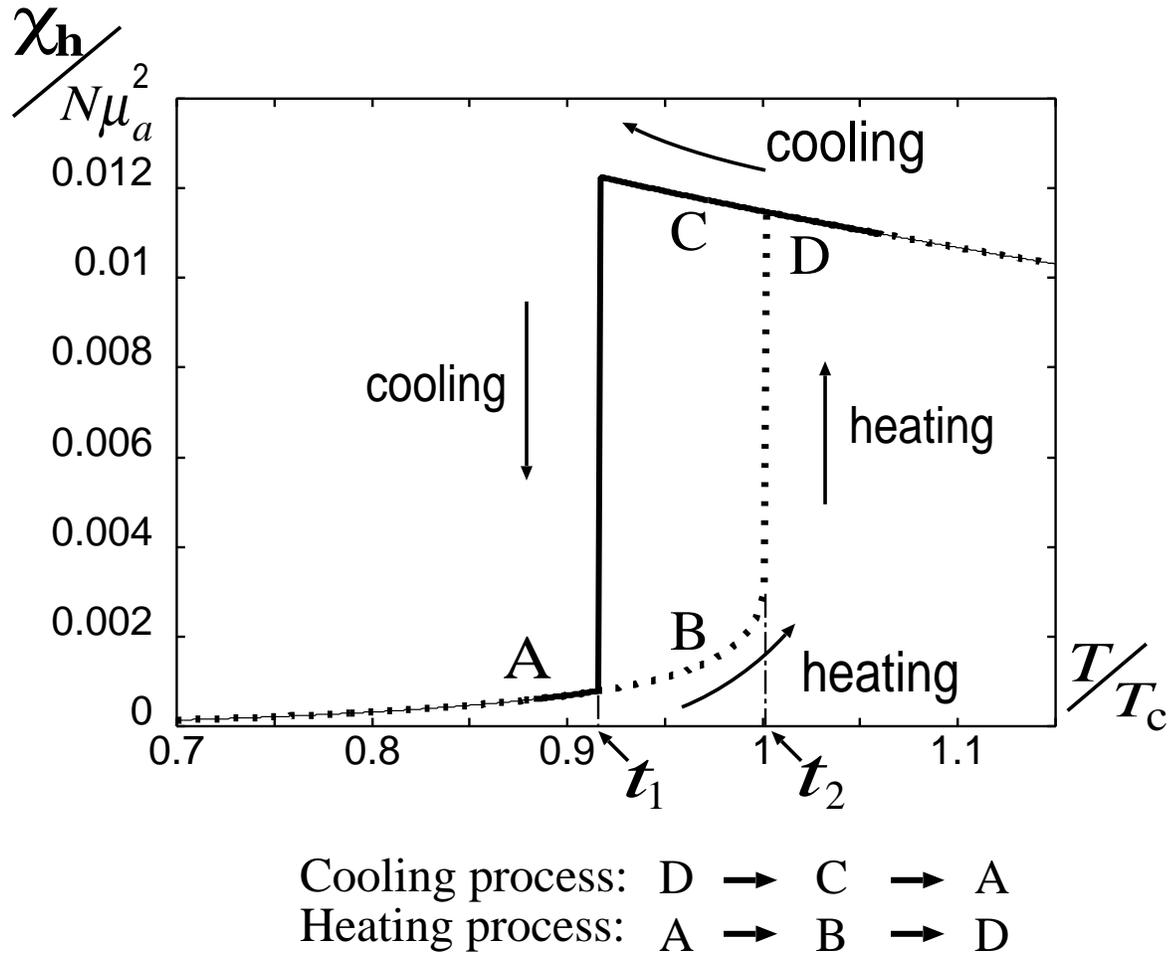,scale=0.8}
\caption{Hysteresis phenomena of uniform susceptibility $\chi_{\rm h}$ versus temperature $T$ in the case of $\varepsilon_{1}=200k_{\rm B}$. The solid line corresponds to cooling process and the dotted line corresponds to heating process with $T_{1}=0.9161T_{c},T_{2}=1.0018T_{c}$.}
\label{unieqhys}
\end{figure}

\begin{figure}
\epsfile{file=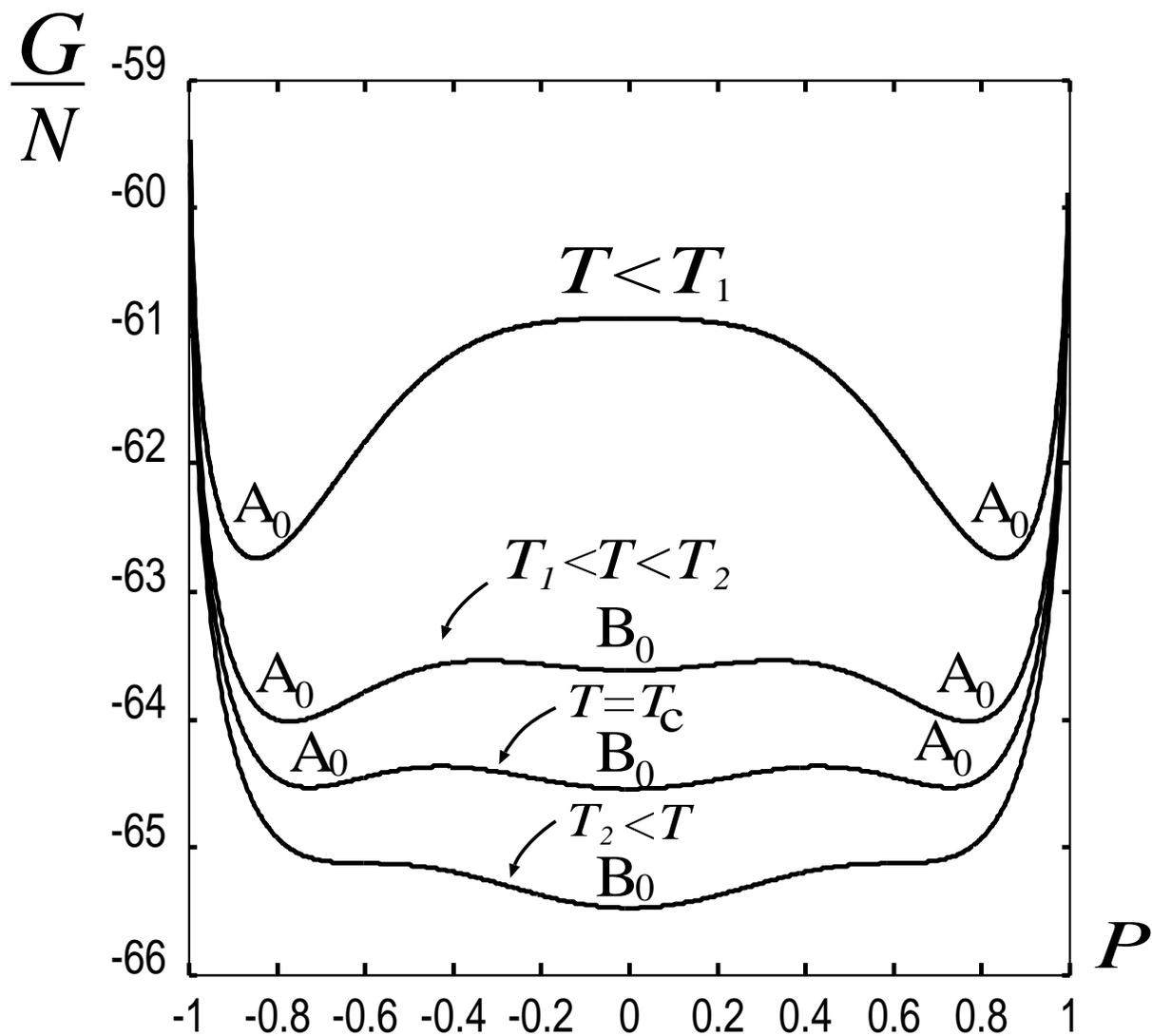,scale=0.8}
\caption{Free energy profile. When $T<T_{1}$ there are two minima at $\rm A_{0}$, $T_{1}<T<T_{2}$ three minima at ${\rm A_{0}, B_{0}}$, $T_{2}<T$ one minimum at $\rm B_{0}$ with $\varepsilon_{1}=200k_{\rm B}$.}
\label{procompare}
\end{figure}

 However, the present uniform susceptibility in the first order transition is completely different from  experimentally observed one.\cite{Kaia} The 
observed  susceptibility shows the hysteresis phenomena versus temperature. Until now, we have 
obtained the susceptibility versus temperature by determining the global  
minimum of the variational free energy at thermal equilibrium for each 
temperature. However, since the heating and cooling process are done at the rate of finite time in the experiment,  we cannot observe the susceptibility 
at thermal equilibrium. In order to solve this discrepancy, we apply the 
iteration method called the NIM (natural iteration method)\cite{Kiku2} for
 numerical calculations of the CVM. The present iteration method leads us to the local minimum of the variational free energy depending upon the initial condition. It should be noted that the local minimum is not always the global minimum of the free energy at each temperature.  With use of the local minimum under the previous temperature  as a starting point the sweeping of temperature continuously gives the hysteresis curve in the heating and the cooling process (Fig. \ref{unieqhys}).  In order to explain the hysteresis phenomena versus temperature we  show  the temperature change in the variational free energy in Fig. \ref{procompare}. Let two temperatures  
at which a drastic change of uniform susceptibility occurs in Fig. \ref{unieqhys} be defined as $T_{1}$ and $T_{2}$ ($T_{1}<T_{c}<T_{2}$). In the heating process, 
(1) when $T<T_{1}$, A in Fig. \ref{unieqhys} exists in a local minimum 
denoted by $\rm A_0$ in Fig. \ref{procompare}, (2) when $T_{1}<T<T_{2}$, B in 
Fig. \ref{unieqhys} exists still in $A_0$ though another local minimum denoted by $\rm B_0$ appears in Fig. \ref{procompare}, (3) at $T=T_{c}$ the three local minima have the same value  and (4) at $T=T_{2}$, B in Fig. \ref{unieqhys} jumps up to D in Fig. \ref{unieqhys} because the state at the unstable $\rm A_{0}$ in Fig. \ref{procompare} tumbles into the stable $\rm B_0$. In the cooling process from D the similar free energy change occurs as a reversible process of the above.
The actual usage of the NIM is explained more in detail in 
Appendix B. \par
\begin{figure}
\epsfile{file=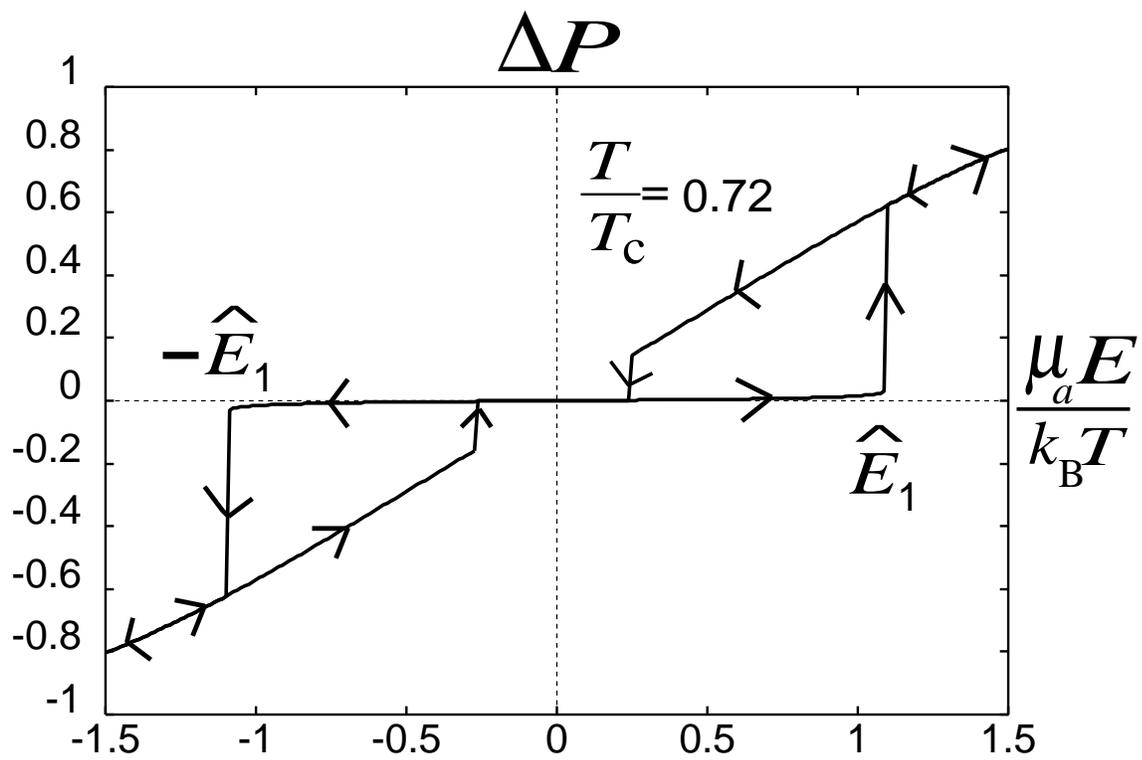,scale=0.8}
\caption{Net polarization $\Delta P$ versus uniform electric field $E$ with $\varepsilon_{1}=200k_{\rm B}$.}
\label{PE}
\end{figure}

 
 We also calculate the hysteresis curve for the net polarization $\Delta P= (P-
P')/2$ versus homogeneous electric field. The numerical result of $\Delta P-E$ 
curve is shown in Fig. {\ref{PE}}.  We can see clearly the transition from the 
anti-ferroelectric state to the polar state at which each polarization in two 
sublattices points to the same direction at $E=E_{1}$ or $E=-E_{1}$.

\section{A Summary and Discussions}
\ \ \ We applied the cactus approximation of the CVM to the Ishibashi model for  the hydrogen bonded ADP-type crystal. The properties of the ADP-type crystal in 
external electric fields were intensively investigated. After re-deriving the 
variational free energy for the Ishibashi model, the equation determining the 
polarization and the susceptibility in a staggered electric field were studied to find the order of the transition. The energy parameter $\varepsilon_1$ characteristic of HPO$_4$ and H$_3$PO$_4$ determines the properties of transition, 
though  $\rm ADP$ undergoes the first order paraelectric-antiferroelectric phase transition in experiments.
We also calculated 
the susceptibility to a homogeneous electric field at thermal equilibrium. 
The calculated susceptibility  does not show any hysteresis even in the parameter 
region of the first order transition, while the homogeneous susceptibility 
observed in experiments shows the hysteresis phenomena in the heating and the cooling process. In order to overcome this 
discrepancy the homogeneous susceptibility in the local minimum of free energy  was calculated and the result is in qualitatively good agreement with the 
experiments. The hysteresis curve is well explained by utilizing the local 
minimum in the variational free energy. Further, though the typical hysteresis curve of the net polarization versus homogeneous 
electric field have not been found in the ADP experiments to our scarce knowledge, we also calculated the hysteresis curve 
depending upon the external electric field with the same idea. The results are 
the ones expected from the Landau theory of the phase transition in the external field.\par 
Though in the Ishibashi model the dipole-dipole interaction is included to induce an anti-ferroelectric transition, the present model without dipole-dipole interaction has essentially the geometrical frustration and is similar to 
the spin ice system with ice rule.  We will discuss the spin ice system from the same view point of the cluster variation method (CVM) in the near future.\par
   The calculation of the dynamical susceptibility for the ADP-type crystal by the dynamical 
cluster variation method\cite{Kiku3} is also in 
progress in order to compare with the experimental data and that\cite{Wada3} for KDP.

\section*{Acknowledgments}
We would like to thank Prof. M. Tokunaga and also our laboratory members of statistical physics for discussions and encouragements.
\appendix
\section{Calculation of Uniform Susceptibility}
From the minimum conditions for the free energy $G$ in eq.(\ref{hfree})\\ $\frac{\partial G}{\partial a_{+}}=\frac{\partial G}{\partial a_{-}}=\frac{\partial G}{\partial d_{+}}=\frac{\partial G}{\partial d_{-}}=\frac{\partial G}{\partial a^{\prime}_{+}}=\frac{\partial G}{\partial a^{\prime}_{-}}=\frac{\partial G}{\partial d^{\prime}_{+}}=\frac{\partial G}{\partial d^{\prime}_{-}}=\frac{\partial G}{\partial a_{0}}=\frac{\partial G}{\partial c_{0}}=\frac{\partial G}{\partial c_{2}}=0,$ 
the following equations are obtained:
\begin{eqnarray}
a_0&=&\frac{1}{(2+2\eta_0+2\eta_2)+tR_1R_2},\nonumber\\
c_{0}&=&\eta_{0}a_{0},\nonumber\\
c_{2}&=&\eta_{2}a_{0},\nonumber\\
a_{+}&=&\frac{R_1}{R_2}ta_{0}\left(p_1r_1e^{2D(p_3-p_4)}\right)h^2,\nonumber\\
a_{-}&=&\frac{R_1}{R_2}ta_{0}\left(p_2r_3e^{-2D(p_3-p_4)}\right)h^{-2},\nonumber
\\       
d_{+}&=&\frac{R_1}{R_2}ta_{0}\eta_1\left(p_1^3p_2r_1^3r_3\right)^{\frac{1}{4}}
e^{D(p_3-p_4)}h,\label{NIM}\\
d_{-}&=&\frac{R_1}{R_2}ta_{0}\eta_1\left(p_1p_2^3r_1r_3^3\right)^{\frac{1}{4}}
e^{-D(p_3-p_4)}h^{-1},\nonumber\\ 
a'_{+}&=&\frac{R_2}{R_1}ta_{0}\left(p_3r_1e^{2D(p_1-p_2)}\right)h^{-2},
\nonumber\\
a'_{-}&=&\frac{R_2}{R_1}ta_{0}\left(p_4r_3e^{-2D(p_1-p_2)}\right)h^2,\nonumber\\       
d'_{+}&=&\frac{R_2}{R_1}ta_{0}\eta_1\left(p_3^3p_4r_1^3r_3\right)^{\frac{1}{4}}
e^{D(p_1-p_2)}h^{-1},\nonumber\\ 
d'_{-}&=&\frac{R_2}{R_1}ta_{0}\eta_1\left(p_3p_4^3r_1r_3^3\right)^{\frac{1}{4}}
e^{-D(p_1-p_2)}h,\nonumber
\end{eqnarray}
where $t, R_1$ and $R_2$ are defined by
\begin{eqnarray}
t&=&\frac{1}{[p_1p_2p_3p_4(r_1r_3)^2]^{\frac{1}{4}}},\nonumber\\
R_1&=&[p_3r_1e^{2D(p_1-p_2)}h^{-2}+p_4r_3e^{-2D(p_1-p_2)}h^{2}\nonumber\\
&+&4\eta_1\left(p_3^3p_4r_1^3r_3\right)^{\frac{1}{4}}
e^{D(p_1-p_2)}h^{-1}\nonumber\\&+&\left(p_3p_4^3r_1r_3^3\right)^{\frac{1}{4}}
e^{-D(p_1-p_2)}h]^{\frac{1}{2}},\nonumber\\
\end{eqnarray}
A set of self-consistent equation for  sublattice polarizations $P$ and $P'$ are obtained from 
\begin{eqnarray}
P&=&p_1-p_2 =a_{+} -a_{-} +2(d_{+} -d_{-}),\nonumber\\
P'&=&p_3-p_4 =a'_{+} -a'_{-} +2(d'_{+} -d'_{-}).\label{hself}
\end{eqnarray}
Substitutions of above relations into eq.(\ref{hself}) lead us to 
\begin{eqnarray}
P&=&\frac{R_1}{R_2}a_0t\left\{p_1r_1e^{2D(p_3-p_4)}h^{2}-p_2r_3e^{-2D(p_3-p_4)}h^{-2}\right.\nonumber\\
&+&
\left. 2\eta_1\bigl[\left(p_1^3p_2r_1^3r_3\right)^{\frac{1}{4}}
e^{D(p_3-p_4)}h\right.\nonumber\\
&-&\left.\left(p_1p_2^3r_1r_3^3\right)^{\frac{1}{4}}
e^{-D(p_3-p_4)}h^{-1}\bigr]\right\},\label{PP'self}\\
P'&=&\frac{R_2}{R_1}a_0t\left\{p_3r_1e^{2D(p_1-p_2)}h^{-2}-p_4r_3e^{-2D(p_1-p_2)}
h^{2}\right.\nonumber\\
&+&\left.2\eta_1\bigl[\left(p_3^3p_4r_1^3r_3\right)^{\frac{1}{4}}
e^{D(p_1-p_2)}h^{-1}\right.\nonumber\\&-&\left.\left(p_3p_4^3r_1r_3^3\right)^{\frac{1}{4}}
e^{-D(p_1-p_2)}h\bigr]\right\},\nonumber
\end{eqnarray}
where it should be noted that $p_1, p_2,  p_3, p_4, r_1, r_3$ are expressed in 
the  sublattice polarization $P$ and $P'$:
\begin{eqnarray}
p_1 &=&\frac{1}{2}(1+P),\quad\quad p_2=\frac{1}{2}(1-P), \nonumber\\
p_3 &=&\frac{1}{2}(1+P'),\quad\quad p_4=\frac{1}{2}(1-P'), \label{p1p3}\\
r_1 &=&\frac{1}{2}(1+\frac{P+P'}{2}),\quad\quad r_3=\frac{1}{2}(1-\frac{P+P'}{2}). \nonumber
\end{eqnarray}
Thus, eq.(\ref{PP'self}) is the self-consistent equation for $P$ and $P'$. Since the linear response of sublattice polarization to the uniform external field $E$ is defined as $\Delta P =P-P_0 $ and $\Delta P'=P'-P_0$,   
the uniform susceptibility $\chi_{\rm h}$ is obtained from eq.({\ref{PP'self}}) as 
\begin{eqnarray}
\chi_{\rm h} &=&\lim_{E\to 0}\frac{N}{2}\mu_a^2\frac{\Delta P -\Delta P'}{E}
=\frac{N\mu^2_{a}}{k_{\rm B}T}\frac{Q_{1}}{Q_{2}},\label{unisus}
\end{eqnarray}
where
\begin{eqnarray}
 Q_{1}&=&4\left[A_{P_0}+{A_{P_0}}^{-1}+\eta_{1}({A_{P_0}}^{\frac{1}{2}}
+{A_{P_0}}^{-\frac{1}{2}})\right]\nonumber\\
&-&\frac{4{P_0}^2}{{\hat{a}_0}^2[A_{P_0}+A_{P_0}^{-1}
+4\eta_{1}({A_{P_0}}^{\frac{1}{2}}
+{A_{P_0}}^{-\frac{1}{2}})]},\nonumber
\end{eqnarray}
\nonumber\\
\section{Natural Iteration Method}
The natural iteration method (NIM)\cite{Kiku2} is one of the method for determining state 
variables from the minimum condition of the variational free energy. Here the 
NIM is briefly explained from starting with eq.($\ref{NIM}$). It should be 
noted that bond variables are characteristic of one proton configuration while PO$_4$ cluster variables are characteristic of four protons configuration. Four protons state variables in eq.($\ref{NIM}$) the left hand side are expressed in terms of only bond state variables in the right hand side.  Let a temperature $T$ and an external field $E$ 
with  energy parameters $\varepsilon_0$, $\varepsilon_1$, $\varepsilon_2$ and $\lambda$ be given. Further, we note that the arbitrary given values of polarization 
$P$ and $P'$ are equivalent to giving the values of bond state variables through eq.(\ref{p1p3}). When the bond state 
variables are substituted into the right hand side of eq.($\ref{NIM}$), the values of four 
proton state variables are naturally calculated.
On the contrary, the geometrical relations under the uniform electric field
\begin{eqnarray}
p_1&=&q_2=c_0+c_2+a_0+a_{+}+3d_{+}+d_{-},\nonumber\\ 
p_2&=&q_1=c_0+c_2+a_0+a_{-}+d_{+}+3d_{-},\nonumber\\
p_3&=&q_4=c_0+c_2+a_0+a'_{+}+3d'_{+}+d'_{-},\nonumber\\
p_4&=&q_3=c_0+c_2+a_0+a'_{-}+d'_{+}+3d'_{-},\nonumber\\
2r_1&=&2s_3=2r_2=2s_4\\
&&\hspace{-1cm}=2c_0+2c_2+2a_0+a_{+}+a'_{+}+3(d_{+}+d'_{+})+(d_{-}+d'_{-}),\label{ite1}\nonumber\\
2r_3&=&2s_1=2r_4=2s_2\nonumber\\
&&\hspace{-1cm}=2c_0+2c_2+2a_0+a_{-}+a'_{-}+(d_{+}+d'_{+})+3(d_{-}+d'_{-})\nonumber
\end{eqnarray}
give values of bond variables. Thus, these bond variables again determine polarizations $P$ and $P'$. Accordingly, this cycle can be repeated until the convergence within some accuracy is reached. Actually it is rigorously proved that as the iteration proceeds, the free energy is always decreased toward a local minimum which is not always the global minimum of the free energy. This property can be fully utilized in problems of hysteresis phenomena.

\newpage
\newpage
\pagebreak
\pagebreak
\newpage
\pagebreak
\newpage
\end{document}